\documentclass[intlimits,twoside,a4paper]{article}

\usepackage[cp1251]{inputenc}

\usepackage[eqsecnum]{cmpj3}

\usepackage{bm}

\issue{2019}{22}{2}{23701}
\doinumber{10.5488/CMP.22.23701}

\title[Laughlin-Jain microscopic wave function approaches]%
{Analytic results of the excited electronic states at $\upsilon=1/3$ and the Laughlin-Jain microscopic wave function approaches%
}
\author[M.A. Ammar]{M.A. Ammar\thanks{E-mail: ahmedammar.mohamed@univ-medea.dz}}
\address{
 Laboratoire de Physique des Techniques Experimentales et ses Applications, University of Medea, Algeria}

\date{Received January 10, 2019, in final form March 17, 2019}

\begin{document}

\maketitle

\begin{abstract}
In this work we studied the properties of a two-dimensional electronic gas
subjected to a strong magnetic field and cooled at a low temperature. We
reported exact analytical results of energies at the ground state. The
results are for systems up to $N_{e}=10$ electrons calculated in the
integer quantum Hall effect (IQHE) regime at the filling factor $%
\upsilon=1 $. To accomplish the calculation we used the complex polar
coordinates method. Note that the system of electrons in the quantum Hall
regime relied heavily on the disk geometry for finite systems of electrons
with arbitrary values of $N_{e}=2$  to $10$  particles. The results that we
obtained by analytical calculations are in good agreement with those
reported by Ciftja [Ciftja O., J. Math. Phys., 2011, \textbf{52}, 122105], where the representation for certain integrals of products of Bessel functions is obtained. In the end, we have studied the composite fermions energies for the excited states for several systems at $\upsilon =1/3$  and the
correspondence between the fractional quantum Hall effect (FQHE) and the IQHE.


\pacs 73.43.-f
\end{abstract}

\section{Introduction}

A discovery of the integer quantum Hall effect (IQHE) \cite{Klitzing80,Tsui82} was the beginning of a big revolution in the field of
condensed matter. It is interesting to study theoretically and numerically
the phases of the integer and fractional quantum Hall effect (FQHE)
in a two-dimensional geometry in order to consider various aspects of
this problem. In this article, we give some details of the formalism used to
treat the IQHE and FQHE problem on the disk geometry \cite{Morf86,Fano87,Ciftja03}. As an hypothesis, Laughlin (1983) \cite{Laughlin83}
considers that the electrons are confined on the plane in a central
symmetric potential, so that the Hall droplet forms a disk of uniform
density in the volume with the correct density of states per unit area. This
area $2\piup ml_{0}^{2}$ is a circle where the radius contains $m$ flux quanta.

We solved the problem of the motion of a confined two-dimensional electron
in a uniform magnetic field. This field is perpendicular to the motion of electrons. That presumes that the electron system is fully spin polarized. The shape
of the system is a disk. Let us first consider free electrons, for a
homogeneous uniform magnetic field $\mathbf{B}=(0,0,B)$. The symmetric gauge
is defined by the vector potential $\mathbf{A}$
\begin{equation}
\mathbf{A}=\frac{\mathbf{B}}{2}(-y,x,0)\text{.}  \label{1}
\end{equation}

In this gauge, the vector potential $\mathbf{A}$ is invariant by rotation
about the axis $z$, and in the canonical momentum $\mathbf{p}=- \text{i} \hbar \pmb\nabla$. In the presence of a magnetic field, the Hamiltonian can be written as follows:
\begin{equation}
\mathbf{H}=\frac{1}{2m}\Big(\mathbf{p}+\frac{e_{0}}{c}\mathbf{A}\Big)^{2}\text{.} 
\label{5}
\end{equation}

By introducing the complex variables
\begin{equation}
z=\frac{x-\text{i}y}{l_{0}}\,,~~\ ~\ ~z^{\ast }=\frac{x+\text{i}y}{l_{0}}\,.  \label{2}
\end{equation}

The characteristic magnetic length on the disk $l_{0}=\sqrt{\hbar/eB}$ that will be taken equal to 1. The wave function in the lowest Landau
level ($n=0$) is denoted by $\Psi _{0,m}$
\begin{equation}
\Psi _{0,m}=\prod_{i<j}\left( z_{i}-z_{j}\right)^{m}\re^{-\sum_{i}|z_{i}|^{2}/4}%
\text{.}  \label{3}
\end{equation}

Let us consider a system of finite size such as a disk of radius $R$, and we can
count the number of states in the lowest Landau level included in this disk.
The probability of presence of the state $\left\vert 0,m\right\rangle $
is defined by
\begin{equation}
\left\vert \Psi _{0,m}(z,\overline{z})\right\vert ^{2}=\frac{1}{2\piup
2^{m}l_{0}^{2}m!}\frac{r^{2m}}{l_{0}^{2m}}\re^{-\frac{r^{2}}{2l_{0}^{2}}}\text{%
.}  \label{4}
\end{equation}

The wave functions in the lowest Landau level are simple monomials in $z$.
Thus, any state in the lowest Landau level (LLL) is given by a
polynomial equation dependent only on $z$. The probability of presence of
the state $\left\vert 0,m\right\rangle $ is maximal over a circle of radius $%
r_{m}$ = $\sqrt{2m}l_{0}$, such that the radial extension of the wave
function is of the order of $l_{0}$. When $m$ increases, the particle moves
symmetrically away from the origin. The last state inside the disk of radius 
$R$ corresponds to $m=R^{2}/2l_{0}^{2}$\ electronic orbitals, which
is also equal to the total number of states in this disk.

The very good approximation of the true fundamental of equation (harmonic
oscillator) of our system is that of Laughlin, but is not exactly the wave
function of this fundamental which is written for odd integer $m$ in equation (\ref%
{3}). This wave function describes a filling state. The basic state used in
our system is described by Laughlin~\cite{Laughlin83} in terms of Slater
determinant in the first quantification which could be used to guess a test wave
function for the ground state of the fractional quantum Hall effect. We
proceed to study the systems $N_e =10$ of the composite fermions (%
CF) wave functions and we have to compare them to the exact wave
functions in the disk geometry.

This article deals with the physics of the path connecting the fractional
quantum Hall effect to the integral quantum Hall effect. In section \ref{sec2}, the model of interaction as well as the Coulomb interaction are presented. In section~\ref{sec3}, the method of analytical results for IQHE at $\upsilon =1$ is shown. In section~\ref{sec4}, we study the excitations of the CF
states. In section~\ref{sec5}, we give the results of our calculus. The conclusion is
presented in section~\ref{sec6}.

\section{Model of interaction}\label{sec2}
The many-electron system is described by the Hamiltonian
\begin{equation}
\widehat{H}=\widehat{K}+\widehat{V}\text{,} 
\end{equation}
where $\widehat{K}$\ is the kinetic energy operator, $\hbar $\ is reduced
Planck's constant, $\omega =e_{0}B/m$\ is the cyclotron
frequency, and the Coulomb interaction $\widehat{V}$\ projected in the 
LLL is obtained starting from the electron-electron interaction,
electron-background and the background-background potentials. When all electrons are confined in the LLL, their kinetic energy is then constant $\big\langle \frac{\widehat{K}}{N_{e}}%
\big\rangle =\frac{1}{2}\hbar \omega $~\cite{Jain07}.
 We consider $N_{e}=10$ electrons of charge ($-e_{0}$) embedded in a uniform
neutralizing background disk of an area and a positive charge $%
N_{e}~e_{0}$. Moreover, the disk is a part of the $xy$ plane subjected to a
strong uniform magnetic field, in the $z$ direction, $\mathbf{B}=Be_{z}$,
and lower temperatures.

The total potential energy operator is defined by
\begin{equation}
\widehat{V}=\widehat{V}_\text{ee}+\widehat{V}_\text{eb}+\widehat{V}_\text{bb}\,,
\label{6}
\end{equation}
with $\widehat{V}_\text{ee}$, $\widehat{V}_\text{eb}$ and $\widehat{V}_\text{bb}$ denoting
the electron-electron, electron-background and the background-background
interaction potentials, respectively. Their corresponding expressions are
given by
\begin{equation}
\widehat{V}_\text{ee}=\sum_{i<j}^{N}\ \frac{e_{0}^{2}}{|r_{i}-r_{j}|}\,,
\label{7}
\end{equation}
\begin{equation}
\widehat{V}_\text{eb}=-\rho \sum_{i=1}^{N}\int\limits_{S_{N}}\rd^{2}r\ \frac{
e_{0}^{2}}{|r_{i}-r|}\,,  \label{8}
\end{equation}
\begin{equation}
\widehat{V}_\text{bb}=\frac{\rho^{2}}{2}\int\limits_{S_{N}}\rd^{2}r\int
\limits_{S_{N}}\rd^{2}r^{\prime }\ \frac{e_{0}^{2}}{|r-r^{\prime }|}\,,
\label{9}
\end{equation}
where $r_{i}$ (or $r_{j}$ ) indicate the electron vector position while $r$
and $r^{\prime }$ are background coordinates. $S_{N}$\ is the area of the
disk and $\rho $\ is the density of the system (the number of electrons per
unit area) that can also be defined by
\begin{equation}
\rho =\frac{\upsilon }{2\piup l_{0}^{2}}\,.  \label{10}
\end{equation}

The integer quantum Hall effect is perfectly explained without invoking interactions: only non-interacting particles fully occupying Landau levels. The interaction is crucial for fractional quantum Hall effect. The
many-body wave functions will be of the form of the equation (\ref{3}). The
theory of the CF is a generalization of those considered by Jain~\cite{Jain89} with the found states at $\upsilon =p/(2pm+1)$ with
integer $m$ and $p$. This theory of the composite fermions immediately
describes the elementary collective excitations by the promotion of a CF 
towards the states in higher Landau levels~\cite{Morf02}.
\begin{equation}
\Psi^\text{gs}(z_{1},\dots ,z_{N_{e}})=\mathcal{P}_\text{LLL}\ \prod_{j<k}\left(
z_{j}-z_{k}\right) ^{2s}\Phi_{n}^\text{gs}.  \label{11}
\end{equation}

Here, $\Phi_{n}^\text{gs}$ represents the incompressible IQHE ground state at
filling factor $\upsilon =1$ (for noninteracting electrons), and $\mathcal{P}_\text{LLL}$ is an operator that
projects the state onto the LLL. The factor $\prod_{j<k}\left(z_{j}-z_{k}\right) ^{2s}$ is the Jastrow factor, binds $2s$ vortices to each electron to convert it into a composite fermion (CF), for two vortices $s=1$~\cite{Jain07}. The method for LLL projection is given in appendix~\ref{appendix_A}. 
\begin{equation}
\Phi _{n}^\text{gs}=\prod_{j<k}^{N_{e}}\left( z_{j}-z_{k}\right) \exp \Big(-\frac{1}{4}\sum_{i}^{N_{e}}|z_{i}|^{2}\Big).   \label{12}
\end{equation}

This can be compared directly with exact numerical diagonalization results
to test the applicability of CF trial wave functions to the system in each
sector with a given number of particles $N_{e}$ and angular momentum $L=mN_{e}(N_{e}-1)/2$, with the basis formed by antisymmetrized functions of
the form (\ref{3}). The background-background interaction potential $\widehat{V}_\text{bb}$\ can be classically calculated without using the wave
function of the electron system. Its value is simply determined by
calculating the elementary defined integral (\ref{9}) and is given by reference~\cite{Ciftja09}. 
\begin{equation}
\big\langle \widehat{V}_\text{bb}\big\rangle=\frac{8N_{e}}{3\piup }\sqrt{\frac{\upsilon N_{e}}{2}}\frac{
e_{0}^{2}}{l_{0}}\,.  \label{13}
\end{equation}

For a given wave function $\Psi (r_{1},\dots ,r_{N_{e}})$, these energies are
determined using the following formulae
\begin{equation}
\big\langle \widehat{V}_\text{ee}\big\rangle=\frac{\left\langle \Psi
\left\vert V_\text{ee}\right\vert \Psi \right\rangle }{\left\langle \Psi
\left\vert \Psi \right. \right\rangle }\,,\,\,\,\,\,
\big\langle \widehat{V}_\text{eb}\big\rangle=\frac{\left\langle \Psi
\left\vert V_\text{eb}\right\vert \Psi \right\rangle }{\left\langle \Psi
\left\vert \Psi \right. \right\rangle }\text{. }
\end{equation}

The term $\widehat{V}_\text{ee}$ is written as follows:
\begin{equation}
\big\langle \widehat{V}_\text{ee}\big\rangle =\frac{N_{e}(N_{e}-1)}{2}\int
\rd^{2}r_{1}\dots \rd^{2}r_{N_{e}}\ \frac{e_{0}^{2}}{|r_{1}-r_{2}|}\left\vert \Psi
r_{1},\dots,r_{N_{e}}\right\vert ^{2}\text{.}  \label{16}
\end{equation}

Similarly, for the $\widehat{V}_\text{eb}$ interaction, we have

\begin{equation}
\big\langle \widehat{V}_\text{eb}\big\rangle =-\rho N_{e}\int
\rd^{2}r_{1}\dots \rd^{2}r_{N_{e}}\left\vert \Psi r_{1},\dots,r_{N_{e}}\right\vert
^{2}\int\limits_{S_{N}} \rd^{2}r\ \frac{e_{0}^{2}}{|r_{1}-r|}\,,
\label{17}
\end{equation}
with
\begin{equation}
\int\limits_{S_{N}}\rd^{2}r\ \frac{e_{0}^{2}}{|r_{1}-r|}=2\piup
R_{N}\int\limits_{0}^{\infty }\frac{\rd q}{q}J_{1}(q)J_{0}\Big(\frac{q}{R_{N}}
r_{1}\Big),  \label{18}
\end{equation}
where is the $J_{n}(x)$ $n$-th order Bessel functions.
A detailed description of the ground state obtained by an analytical method is given in our earlier
works~\cite{Ammar16,Ammar18}.

\section{Analytical results for IQHE at $\protect\upsilon =1$}\label{sec3}

In this section, we obtain the analytical expressions for the total energy per particle (in
units $e_{0}^{2}/l_{0}$) and related quantities corresponding to IQHE system of electrons in a disk geometry at $\upsilon =1$.
The ground state interaction energy per particle can be written as follows:
\begin{equation}
\varepsilon =\varepsilon_\text{ee}+\varepsilon_\text{eb}+\varepsilon_\text{bb}\,,
\end{equation}
where $\varepsilon =\big\langle \widehat{V}\big\rangle/N_{e}$\,,\,\, $\varepsilon_\text{ee}=\big\langle \widehat{V}_\text{ee}\big\rangle /N_{e}$\,,\,\, $\varepsilon_\text{eb}=\big\langle \widehat{V}_\text{eb}\big\rangle / N_{e}$ and $\varepsilon _\text{bb}=\big\langle \widehat{V}_\text{bb}\big\rangle/N_{e}$. The energy of the interaction between electrons-electrons, electrons-background and the background-background is given by

$N_{e}=2;\left\{ 
\begin{array}{l}
\varepsilon _\text{ee}=\frac{\sqrt{\piup }}{8}\frac{e_{0}^{2}}{l_{0}}=0.22155673%
\frac{e_{0}^{2}}{l_{0}} \\ 
\varepsilon _\text{bb}=0.49007013\frac{e_{0}^{2}}{l_{0}} \\ 
\varepsilon _\text{eb}=-\frac{\sqrt{2 \piup }}{4e}[3I_{0}(1)+5I_{1}(1)]=
-1.52705731\frac{e_{0}^{2}}{l_{0}} \\ 
\varepsilon =-0.45667421\frac{e_{0}^{2}}{l_{0}}
\end{array}
\right. $

$N_{e}=3;\left\{ 
\begin{array}{l}
\varepsilon _\text{ee}=\frac{87\sqrt{\piup }}{384}\frac{e_{0}^{2}}{l_{0}}=0.40157158%
\frac{e_{0}^{2}}{l_{0}} \\ 
\varepsilon _\text{bb}=1.03959573\frac{e_{0}^{2}}{l_{0}} \\ 
\varepsilon _\text{eb}=-\frac{\sqrt{2\piup }}{16}[9I_{0}\big(\frac{3}{2}\big)+41I_{1}\big(\frac{
3}{2}\big)]=-1.92501490\frac{e_{0}^{2}}{l_{0}} \\ 
\varepsilon =-0.48384759\frac{e_{0}^{2}}{l_{0}}%
\end{array}%
\right. $

$N_{e}=4;\left\{ 
\begin{array}{l}
\varepsilon _\text{ee}=\frac{5147\sqrt{\piup }}{16384}\frac{e_{0}^{2}}{l_{0}}%
=0.55681274\frac{e_{0}^{2}}{l_{0}} \\ 
\varepsilon _\text{bb}=1.20042175\frac{e_{0}^{2}}{l_{0}} \\ 
\varepsilon _\text{eb}=\frac{\sqrt{2\piup }}{96}[79I_{0}(2)-515I_{1}(2)]=-2.25835527\frac{e_{0}^{2}}{l_{0}} \\ 
\varepsilon =-0.50112077\frac{e_{0}^{2}}{l_{0}}%
\end{array}%
\right. $

$N_{e}=5;\left\{ 
\begin{array}{l}
\varepsilon _\text{ee}=\frac{102819\sqrt{\piup }}{262144}\frac{e_{0}^{2}}{l_{0}}%
=0.69519780\frac{e_{0}^{2}}{l_{0}} \\ 
\varepsilon _\text{bb}=1.34211232\frac{e_{0}^{2}}{l_{0}} \\ 
\varepsilon _\text{eb}=\frac{5\sqrt{2\piup }}{768e^{5/2}}[851I_{0}\big(\frac{5}{2}%
\big)-1869I_{1}\big(\frac{5}{2}\big)]=-2.55064421\frac{e_{0}^{2}}{l_{0}}
\\ 
\varepsilon =-0.51333409\frac{e_{0}^{2}}{l_{0}}%
\end{array}%
\right. $

$N_{e}=6;\left\{ 
\begin{array}{l}
\varepsilon _\text{ee}=\frac{15545879\sqrt{\piup }}{33554432}\frac{e_{0}^{2}}{l_{0}}%
=0.82118371\frac{e_{0}^{2}}{l_{0}} \\ 
\varepsilon _\text{bb}=1.47021039\frac{e_{0}^{2}}{l_{0}} \\ 
\varepsilon _\text{eb}=\frac{3\sqrt{2\piup }}{2560e^{3}}
[16629I_{0}(3)-25397I_{1}(3)]=-2.81395015\frac{e_{0}^{2}}{
l_{0}} \\ 
\varepsilon =-0.52255605\frac{e_{0}^{2}}{l_{0}}
\end{array}
\right. $

$N_{e}=7;\left\{ 
\begin{array}{l}
\varepsilon _\text{ee}=\frac{283985723\sqrt{\piup }}{536870912}\frac{e_{0}^{2}}{
l_{0}}=0.93756539\frac{e_{0}^{2}}{l_{0}} \\ 
\varepsilon _\text{bb}=1.58800872\frac{e_{0}^{2}}{l_{0}} \\ 
\varepsilon _\text{eb}=\frac{7\sqrt{2\piup }}{92160~e^{7/2}}[775265~I_{0}\big(\frac{7}{2}\big)-1007359~I_{1}\big(\frac{7}{2}\big)]=-3.05541141\frac{e_{0}^{2}}{l_{0}} \\ 
\varepsilon =-0.52983730\frac{e_{0}^{2}}{l_{0}}
\end{array}
\right. $

$N_{e}=8;\left\{ 
\begin{array}{l}
\varepsilon _\text{ee}=\frac{81126302255\sqrt{\piup }}{137438953472}\frac{e_{0}^{2}%
}{l_{0}}=1.04622906\frac{e_{0}^{2}}{l_{0}} \\ 
\varepsilon _\text{bb}=1.69765273\frac{e_{0}^{2}}{l_{0}} \\ 
\varepsilon _\text{eb}=\frac{\sqrt{2\piup }}{1290240~e^{4}}%
[217069951~I_{0}(4)-260821371~I_{1}(\frac{7}{2})]=-3.27965711\frac{e_{0}^{2}}{l_{0}} \\ 
\varepsilon =-0.53577533\frac{e_{0}^{2}}{l_{0}}%
\end{array}%
\right. $

$N_{e}=9;\left\{ 
\begin{array}{l}
\varepsilon _\text{ee}=\frac{1424926487975\sqrt{\piup }}{2199023255552}\frac{%
e_{0}^{2}}{l_{0}}=1.14851739\frac{e_{0}^{2}}{l_{0}} \\ 
\varepsilon _\text{bb}=1.80063263\frac{e_{0}^{2}}{l_{0}} \\ 
\varepsilon _\text{eb}=\frac{3\sqrt{2\piup }}{2293760e^{\frac{9}{2}}}%
[358919853~I_{0}\big(\frac{9}{2}\big)-413936659I_{1}\big(\frac{9}{2}\big)]=-3.48988824\frac{e_{0}^{2}}{l_{0}} \\ 
\varepsilon =-0.54073822\frac{e_{0}^{2}}{l_{0}}
\end{array}
\right. $

$N_{e}=10;\left\{ 
\begin{array}{l}
\varepsilon _\text{ee}=\frac{395560272250157\sqrt{\piup }}{562949953421312}\frac{
e_{0}^{2}}{l_{0}}=1.24542568\frac{e_{0}^{2}}{l_{0}} \\ 
\varepsilon _\text{bb}=1.89803345\frac{e_{0}^{2}}{l_{0}} \\ 
\varepsilon _\text{eb}=\frac{5\sqrt{2\piup }}{10616832~e^{5}}[\text{2751387571}
~I_{0}(5)-\text{3098794619}I_{1}(5)]=-3.68842587\frac{e_{0}^{2}}{l_{0}
} \\ 
\varepsilon =-0.54496674\frac{e_{0}^{2}}{l_{0}}\,.
\end{array}
\right. $

\section{The quasielectron ($qe$) energies}\label{sec4}

Nowadays, there are two universally accepted theories in the field of FQHE,
the theory of Laughlin~\cite{Laughlin83} and the theory of Jain~\cite{Jain89}.
An early trial wave function proposed by Laughlin for the ground state at
the filling factor $\upsilon =1/m $,  $m $ odd, turned out to work well~\cite
{GunSangJeon03}. The CF theory applies to a broader range of phenomena,
while also providing a new interpretation for the physics of the $\upsilon 
= 1/m $ state, as a state of CF at an effective filling of $\upsilon^* = 1 $
\cite{Jain89,GunSangJeon03}. While the wave function for the $\upsilon  =
1/m $ ground state from the CF theory is the same as that in reference~\cite{Laughlin83}, the wave functions for the excitations are different, which
gives an opportunity to test the validity of the CF theory at $\upsilon
= 1/m$ itself. We note that when speaking of ``quasielectrons'' in this paper, we really mean
``quasielectron excitations'' of an
incompressible FQHE state. Our objective in this work is to compare the
two theories for systems containing $(N_{e}-1)qe$.

\subsection{Laughlin's $N_{e}-1$ quasielectron wave function}

We concentrate herein below on $\upsilon $ = 1/3, with one quasielectron in
the disk geometry. Note that the wave function $\Psi _\text{L}^{(N_{e}-1)qe}$\
corresponding to the $(N_{e}-1)$ quasielectron states of Laughlin to the
filling factor $\upsilon  =1/3$. The trick of piercing a flux quantum
adiabatically through the system motivates the following wave functions for
the quasielectron\cite{Laughlin83}
\begin{equation}
\Psi _\text{L}^{(N_{e}-1)qe}=\Psi _{m}^{-z_{0}}=\exp \left(-\sum_{j}\ \frac{|z_{j}|^{2}
}{4}\right)\left[ \prod_{i=1}^{N_{e}-1}\Bigg(\frac{\partial }{\partial z_{i}}-\frac{z_{0}}{
l_{0}^{2}}\Bigg)\right] \Bigg[ \prod_{j<k}\left( z_{j}-z_{k}\right)
^{m}\Bigg]. 
\end{equation}

In physics, this equation describes the creation of $(N_{e}-1)$ quasielectrons located at the origin.

\subsection{The compact states [1,1, \ldots,1]}

\looseness=-1 In this subsection, we have used only one consequence of the CF theory for
the quasiparticules in occupation [1,1, \dots,1] presented in~\cite%
{GunSangJeon03,Jain95,Dev92,GunSangJeon07}. The CF basis consists of Jain
wave function~\cite{Jain97}, where the derivatives do not act on the Gaussian
factor~\cite{Dev92}, and one will derive only the polynomial part of the wave
function from. The composite fermion wave function for the quasiparticle at $%
\upsilon  =1/3$ is then given by

\begin{equation}
{\tiny \Psi }_\text{CF}^{1qe}{\tiny =P}_\text{LLL}{\tiny \ }\prod_{j<k}^{N}\left(
z_{j}-z_{k}\right) ^{2}\left\vert 
\begin{array}{ccccc}
\overline{z}_{1} & \overline{z}_{2} & \cdot & \cdot & \overline{z}_{N_{e}}
\\ 
1 & 1 & \cdot & \cdot & 1 \\ 
z_{1} & z_{2} & \cdot & \cdot & z_{N_{e}} \\ 
\cdot & \cdot & \cdot & \cdot & \cdot \\ 
\cdot & \cdot & \cdot & \cdot & \cdot \\ 
z_{1}^{N_{e}-2} & z_{2}^{N_{e}-2} & \cdot & \cdot & z_{N_{e}}^{N_{e}-2}
\end{array}%
\right\vert { \re}^{-\sum_{j=1}^{N}|z_{j}|^{2}/4},
\label{20}
\end{equation}
and the CF wave function for the $N_{e}-1$ quasiparticle at $\upsilon =
\frac{1}{3}$ is then given by
\begin{equation}
{\tiny \Psi }_\text{CF}^{\left[ 1,...,1\right] }{\tiny =P}_\text{LLL}{\tiny \ }
\prod_{j<k}^{N}\left( z_{j}-z_{k}\right) ^{2}\left\vert 
\begin{array}{cccc}
(\overline{z}_{1})^{N_{e}-1} & (\overline{z}_{2})^{N_{e}-1} & \cdot & (
\overline{z}_{N_{e}})^{N_{e}-1} \\ 
(\overline{z}_{1})^{N_{e}-2} & (\overline{z}_{2})^{N_{e}-2} & \cdot & (
\overline{z}_{N_{e}})^{N_{e}-2} \\ 
\cdot & \cdot & \cdot & \cdot \\ 
(\overline{z}_{1})^{2} & (\overline{z}_{2})^{2} & \cdot & (\overline{z}
_{N_{e}})^{2} \\ 
\overline{z}_{1} & \overline{z}_{2} & \cdot & \overline{z}_{N_{e}} \\ 
1 & 1 & \cdot & 1
\end{array}
\right\vert {\re}^{-\sum_{j=1}^{N}|z_{j}|^{2}/4}.
\label{21}
\end{equation}

We have studied the Laughlin and CF energies for the excited
states for several systems at $\upsilon=1/3$ and the correspondence
between the FQHE and the IQHE. These energies are given in table~\ref{tab:table1}, where $V_\text{L}$ represents the electron-electron interaction energy of Laughlin's wave function, $V_\text{CF}$ is the electron-electron interaction energy of Jain's wave function, and $V_\text{Exact}$ is the electron-electron interaction
energy of exact analytical expressions \cite{Ciftja11}.

\begin{table}[htb]%
\caption{Comparison between the energy of Laughlin wave function and the
energy of the CF wave function in disk geometry. Energies are in units of
$e_0^2/l_0$\,.}\renewcommand{\tabcolsep}{1pc}
\label{tab:table1}
\vspace{2ex}
\begin{tabular}[b]{|c|c|c|c|c|c|c|c|c|}
\hline
$_{^\text{CF}}$ & \multicolumn{2}{|c|}{$_{^{\text{1~qe}\equiv \text[N_{e}
-1,1]}}$} & \multicolumn{2}{|c|}{$_{^{\text{2~qe}\equiv \text[N_{e}
-2,1,1]}}$} & \multicolumn{2}{|c|}{$_{^{\text{3~qe}\equiv \text[N_{e}
-3,1,1,1]}}$} & \multicolumn{2}{|c|}{$_{^{\text{4~qe}\equiv \text{[}N_{e}%
\text{-4,1,1,1,1]}}}$} \\ \cline{2-9}
$_{N_{e}}$ & ${V}_\text{L}$ & ${V}_\text{CF}$ & ${V}_\text{L}$ & ${V}_\text{CF}$ & ${V}_\text{L}$ & ${V}_\text{CF}$ & ${V}_\text{L}$ & ${V}_\text{CF}$ \\ \hline
$_{2}$ & {\scriptsize 0.443114} & {\scriptsize 0.443114} & $_{-}$ & $_{-}$ & 
$_{-}$ & $_{-}$ & $_{-}$ & $_{-}$ \\ 
$_{3}$ & {\scriptsize 0.891204} & {\scriptsize 0.891204} & {\scriptsize %
1.20472} & {\scriptsize 1.20472} & $_{-}$ & $_{-}$ & $_{-}$ & $_{-}$ \\ 
$_{4}$ & {\scriptsize 1.50139} & {\scriptsize 1.50172} & {\scriptsize %
1.78598} & {\scriptsize 1.78512} & {\scriptsize 2.22725} & {\scriptsize %
2.22725} & $_{-}$ & $_{-}$ \\ 
$_{5}$ & {\scriptsize 2.24874} & {\scriptsize 2.24905} & {\scriptsize %
2.53811} & {\scriptsize 2.53707} & {\scriptsize 2.92117} & {\scriptsize %
2.91876} & {\scriptsize 3.47599} & {\scriptsize 3.47599} \\ 
$_{6}$ & {\scriptsize 3.11368} & {\scriptsize 3.11219} & {\scriptsize %
3.42216} & {\scriptsize 3.41913} & {\scriptsize 3.79696} & {\scriptsize %
3.79441} & {\scriptsize 4.26863} & {\scriptsize 4.26452} \\ 
$_{7}$ & {\scriptsize 4.08010} & {\scriptsize 4.07622} & {\scriptsize %
4.41587} & {\scriptsize 4.40773} & {\scriptsize 4.80306} & {\scriptsize %
4.79646} & {\scriptsize 5.25455} & {\scriptsize 5.27973} \\ \hline\hline
$_{^\text{CF}}$ & \multicolumn{2}{|c|}{$_{^{\text{5~qe}\equiv [N_{e}
-5,1,1,1,1,1]}}$} & \multicolumn{2}{|c|}{$_{^{\text{6~qe}\equiv \text{%
[1,1,1,1,1,1,1]}}}$} & \multicolumn{4}{|c|}{$_{^\text{IQHE}}$} \\ \cline{2-9}
$_{N_{e}}$ & ${V}_\text{L}$ & ${V}_\text{CF}$ & ${V}_\text{L}$ & ${V}_\text{CF}$ & \multicolumn{4}{|c|}{${V}_\text{Exact}$\cite{Ciftja11}} \\ \hline
$_{2}$ & $_{-}$ & $_{-}$ & $_{-}$ & $_{-}$ & \multicolumn{4}{|c|}%
{\scriptsize 0.443114} \\ 
$_{3}$ & $_{-}$ & $_{-}$ & $_{-}$ & $_{-}$ & \multicolumn{4}{|c|}%
{\scriptsize 1.20472} \\ 
$_{4}$ & $_{-}$ & $_{-}$ & $_{-}$ & $_{-}$ & \multicolumn{4}{|c|}%
{\scriptsize 2.22725} \\ 
$_{5}$ & $_{-}$ & $_{-}$ & $_{-}$ & $_{-}$ & \multicolumn{4}{|c|}%
{\scriptsize 3.47599} \\ 
$_{6}$ & {\scriptsize 4.92710} & {\scriptsize 4.92710} & $_{-}$ & $_{-}$ & 
\multicolumn{4}{|c|}{\scriptsize 4.92710} \\ 
$_{7}$ & {\scriptsize 5.80798} & {\scriptsize 5.80254} & {\scriptsize %
6.56296} & {\scriptsize 6.56296} & \multicolumn{4}{|c|}{\scriptsize %
6.56296} \\ \hline
\end{tabular}
\end{table}

Our study confirms the description of quasielectrons as (CF) in an excited (CF) quasi-Landau level. This wave
function contains one excited (CF) in the second CF
quasi-Landau level but the other in the $(N_{e}-1)$ CF quasi-Landau level,
as indicated by the notation $[1,1, \dots,1]$.

\section{Results and discussion}\label{sec5}

Within this work, we considered two systems. The first one for the ground
energy with up to $N_{e}=10$ electrons at the filling $\upsilon=1$
and the second one for the exited energy with up to $N_{e}=7$ electrons
at the filling $\upsilon=1/3$. Several researchers used the disk
geometry in their works~\cite{Morf86,Ciftja03,Laughlin83,Ammar16}. The
mathematical derivations as well as the Mathematica code~\cite{Mathematica4}
are used to calculate the interaction energy to perform the analytical
method for $N_{e}=7$ particles. The code of the electron-background interaction energy computation $%
\varepsilon _\text{eb}$ is shown in appendix~\ref{appendix_B}.

\begin{figure}[htb]
\centerline{\includegraphics[width=0.75\textwidth]{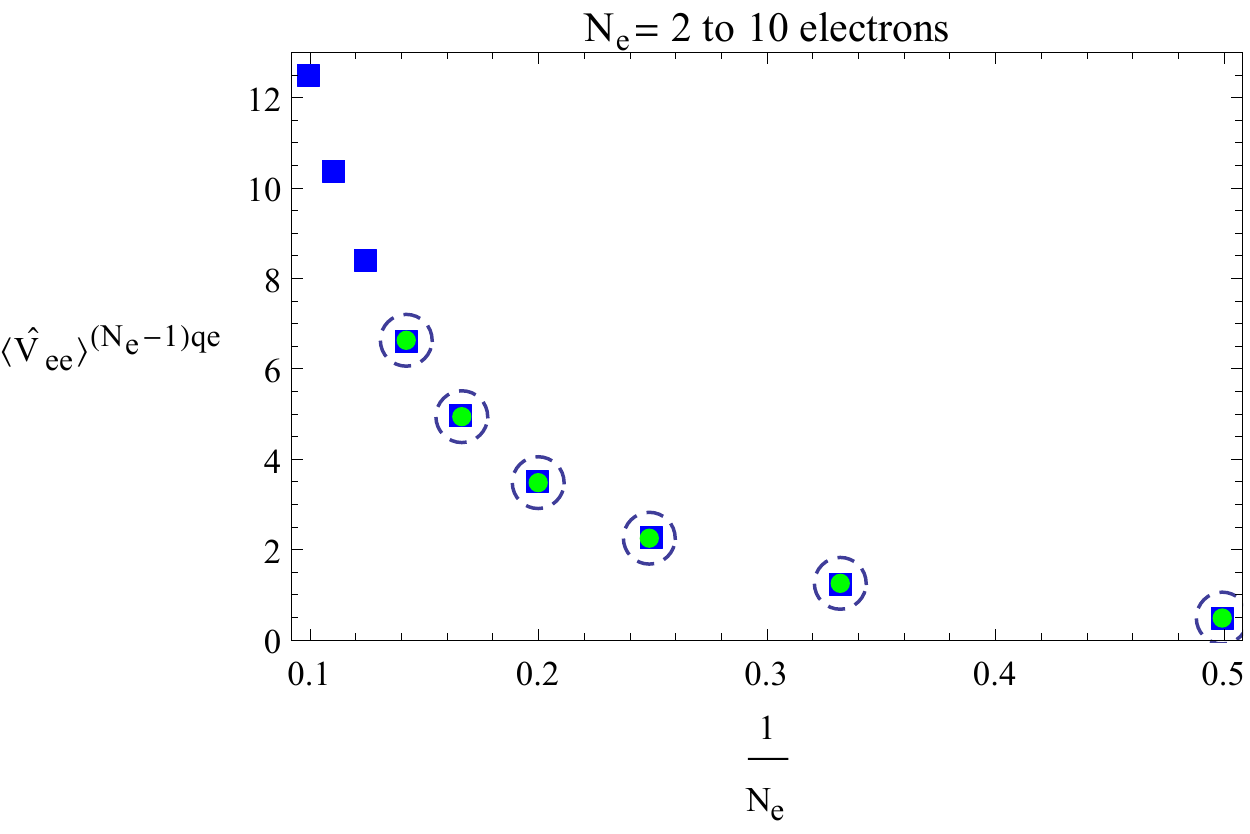}}
\vspace{-3mm}
\caption{(Colour online) The $N_e-1$ quasielectron energies  for $N_e=7$ in disk geometry at $\upsilon=1/3$. The green thickness represents Laughlin's states
$\Psi$$_\text{L}$$\mathbb{}^{(N_e-1)qp}$. The dashed circle represents Jain's states
$\Psi$$_\text{CF}$$\mathbb{}^{[1,1, \ldots\,, 1]}$. The blue rectangle represents IQHE states at $\upsilon=1$.} \label{fig-smp1}
\end{figure}

This result is consistent with previous studies, the Laughlin wave function
for $N_{e}$ quasiparticle is increased and approaches about Jain states in
occupation [1,1, \dots,1]. Figure~\ref{fig-smp1} shows that the energies of $Nqe$ quasielectrons
for $N_{0}=2$ to 7 particles, in the disk geometry for $\upsilon=1/3$,
and the FQHE are equal to those for the IQHE, where we denote the
compact states by [$N_{0}$, $N_{1}$, \dots, $N_{7}$], where one composite
fermion occupied one Landau level~\cite{Jain95,Dev92}.

In figure~\ref{fig-smp1}, we can see that the results derived by the present exact
analytical calculation at the filling $\upsilon=1$ and $\upsilon=1/3$
(the excited states) compare well with the results of~\cite{Ciftja11}
obtained using the method of Exact analytic solutions. For $N_{e}$ quasiparticles, the
energies for $N_{e}=7$ in disk geometry at $\upsilon=1/3$ of Laughlin
states correspond to occupation [1,1, \dots,1] of Jain states.

\section{Conclusion}\label{sec6}

We conclude that the analytical expression for the Coulomb interaction energy
in disk geometry at filling factor 1/3 permits us to obtain a correspondence
between the fractional quantum Hall effect and the integer quantum Hall effect. 
 The coinciding energies of the IQHE and the $(N_e-1)$ quasielectron of the FQHE states are expected, where the FQHE and the IQHE can be unified. The FQHE is explained as the IQHE of composite fermions.
This physics not only gives the best available microscopic wave functions for the quasiparticles but also
brings out new qualitative structures for multi-quasiparticle states. These quasiparticles may be classified in theory as composite fermions, which allows us to understand the fractional quantum hall effect of
electrons as an integer quantum hall effect of these composite
fermions.

\appendix
\section{Lowest Landau level projection operator $\mathcal{P}_\text{LLL}$}
\label{appendix_A}

In the CF theory, the wave function for the ground state $\Psi ^\text{GS}$ at $\upsilon =1/3$ takes the form~\cite{Jain89}\ 
\begin{equation}
\Psi ^\text{GS}(z_{1},\dots ,z_{N_{e}})=\mathcal{P}_\text{LLL}\ \prod_{j<k}\left(
z_{j}-z_{k}\right) ^{2}\Phi _{1}\,,  \label{6.2}
\end{equation}
where $\Phi _{1}$ represents the wave function of noninteracting electrons%
\begin{equation}
\Phi _{1}=\prod_{j<k}^{N}\left( z_{j}-z_{k}\right) \exp \bigg(-\frac{1}{4}%
\sum_{i}^{N}|z_{i}|^{2}\bigg)\text{.}  \label{6.3}
\end{equation}

The CF wave function for one quasielectron (1$qe$) at $\upsilon =1/3$ is
then given by
\begin{equation}
\Psi _\text{CF}^{1qp}=\mathcal{P}_\text{LLL}\ \prod_{j<k}^{N}\left( z_{j}-z_{k}\right)
^{2}\left\vert 
\begin{array}{ccccc}
\overline{z}_{1} & \overline{z}_{2} & \cdot  & \cdot  & \overline{z}_{N} \\ 
1 & 1 & \cdot  & \cdot  & 1 \\ 
z_{1} & z_{2} & \cdot  & \cdot  & z_{N} \\ 
\cdot  & \cdot  & \cdot  & \cdot  & \cdot  \\ 
z_{1}^{N-2} & z_{2}^{N-2} & \cdot  & \cdot  & z_{N}^{N-2}%
\end{array}%
\right\vert \exp \bigg(-\frac{1}{4}\sum_{j}^{N}|z_{j}|^{2}\bigg)\,.  \label{6.4}
\end{equation}

The $\mathcal{P}_\text{LLL}$ is an operator that projects the state onto the
lowest Landau level, where the LLL projection of any wave function can be
obtained by normal ordering the wave function followed by replacing $%
\overline{z}_{i}\rightarrow2\partial /\partial z_{i}$, where the
derivatives do not act on the Gaussian factor.

For example, for $N_{e}$ $=4$ electrons
\begin{equation}
\Psi _\text{CF}^{1qp}=\mathcal{P}_\text{LLL}\ \prod_{j<k}^{N}\left( z_{j}-z_{k}\right)
^{2}\left\vert 
\begin{array}{cccc}
\overline{z}_{1} & \overline{z}_{2} & \overline{z}_{3} & \overline{z}_{4} \\ 
1 & 1 & 1 & 1 \\ 
z_{1} & z_{2} & z_{3} & z_{4} \\ 
z_{1}^{2} & z_{2}^{2} & z_{3}^{2} & z_{4}^{2} \\ 
z_{1}^{3} & z_{2}^{3} & z_{3}^{3} & z_{4}^{3}%
\end{array}%
\right\vert \exp \bigg(-\frac{1}{4}\sum_{j}^{N}|z_{j}|^{2}\bigg)\,,
\end{equation}
where
\begin{eqnarray}
\Phi _{1}^{^{\prime }} &=&\prod_{j<k}^{N}\left( z_{j}-z_{k}\right) ^{2} 
\nonumber \\
&=&(z_{1}-z_{2})^{2}(z_{1}-z_{3})^{2}(z_{2}-z_{3})^{2}(z_{1}-z_{4})^{2}(z_{2}-z_{4})^{2}(z_{3}-z_{4})^{2}%
\text{.}
\end{eqnarray}

The CF basis functions of LLL projected wave functions $\Psi _\text{CF}^{1qp}$, then take the following form: 
\begin{eqnarray}
\Psi _\text{CF}^{1qp} &=&\bigg(z_{2}^{2}z_{3}\frac{\partial \Phi _{1}^{^{\prime }}}{%
\partial z_{1}}-z_{2}z_{3}^{2}\frac{\partial \Phi _{1}^{^{\prime }}}{%
\partial z_{1}}-z_{2}^{2}z_{4}\frac{\partial \Phi _{1}^{^{\prime }}}{%
\partial z_{1}}+z_{3}^{2}z_{4}\frac{\partial \Phi _{1}^{^{\prime }}}{%
\partial z_{1}}+z_{2}z_{4}^{2}\frac{\partial \Phi _{1}^{^{\prime }}}{%
\partial z_{1}}  \nonumber \\
&&-z_{3}z_{4}^{2}\frac{\partial \Phi _{1}^{^{\prime }}}{\partial z_{1}}%
-z_{1}^{2}z_{3}\frac{\partial \Phi _{1}^{^{\prime }}}{\partial z_{2}}%
+z_{1}z_{3}^{2}\frac{\partial \Phi _{1}^{^{\prime }}}{\partial z_{2}}%
+z_{1}^{2}z_{4}\frac{\partial \Phi _{1}^{^{\prime }}}{\partial z_{2}}%
-z_{3}^{2}z_{4}\frac{\partial \Phi _{1}^{^{\prime }}}{\partial z_{2}} 
\nonumber \\
&&-z_{1}z_{4}^{2}\frac{\partial \Phi _{1}^{^{\prime }}}{\partial z_{2}}%
+z_{3}z_{4}^{2}\frac{\partial \Phi _{1}^{^{\prime }}}{\partial z_{2}}%
+z_{1}^{2}z_{2}\frac{\partial \Phi _{1}^{^{\prime }}}{\partial z_{3}}%
-z_{1}z_{2}^{2}\frac{\partial \Phi _{1}^{^{\prime }}}{\partial z_{3}}%
-z_{1}^{2}z_{4}\frac{\partial \Phi _{1}^{^{\prime }}}{\partial z_{3}} 
\nonumber \\
&&+z_{2}^{2}z_{4}\frac{\partial \Phi _{1}^{^{\prime }}}{\partial z_{3}}%
+z_{1}z_{4}^{2}\frac{\partial \Phi _{1}^{^{\prime }}}{\partial z_{3}}%
-z_{2}z_{4}^{2}\frac{\partial \Phi _{1}^{^{\prime }}}{\partial z_{3}}%
-z_{1}^{2}z_{2}\frac{\partial \Phi _{1}^{^{\prime }}}{\partial z_{4}}%
+z_{1}z_{2}^{2}\frac{\partial \Phi _{1}^{^{\prime }}}{\partial z_{4}} 
\nonumber \\
&&+z_{1}^{2}z_{3}\frac{\partial \Phi _{1}^{^{\prime }}}{\partial z_{4}}%
-z_{2}^{2}z_{3}\frac{\partial \Phi _{1}^{^{\prime }}}{\partial z_{4}}%
-z_{1}z_{3}^{2}\frac{\partial \Phi _{1}^{^{\prime }}}{\partial z_{4}}%
+z_{2}z_{3}^{2}\frac{\partial \Phi _{1}^{^{\prime }}}{\partial z_{4}}%
\bigg) \re^{-[(z_{1}^{2}+z_{2}^{2}+z_{3}^{2}+z_{4}^{2})/4]}\,.
\end{eqnarray}


\section{The $\protect\varepsilon _\text{eb}$ calculation}
\label{appendix_B}

\begin{verbatim}
(*SetDirectory["directory name"]*)
SetDirectory["C:\Users\Hmida\Desktop\4-particles-electron-background"];
Directory[];
Module[{Ne, Nu, R, SN, Rho, Polyz, CoefPoly, CoefPolyMin, PolyExpand,
RePoly, CoefPolyList, Clist, Inner1, Inlist, I1, I2, I3, I5, I6,
DenomiIntegralr1, SumDenomi, SumNomi, Ke, KeN, KeM, CoefPoly1, SumGloVeb, 
EnergyVeb, strm}, Ne = 4;Nu = 1;R = l[0]*Sqrt[2*Ne/Nu];SN = Pi*R^2;
Rho = Nu/(2*Pi*l[0]^2);
Polyz = Product[Product[(z[i] - z[j]), {j, i + 1, Ne}], {i, 1, Ne - 1}];
CoefPoly = Exponent[Polyz, z[1]];
CoefPolyMin = Exponent[Polyz, z[1], Min];
PolyE = Expand[Polyz];
RePoly = Flatten[Table[ Coefficient[PolyE, z[1], i], {i, CoefPolyMin,CoefPoly}]];
CoefPolyList = Plus @@ Select[RePoly, #1 =!= 0 &];
Clist = CoefPolyList /. Plus -> List;
Inner1 = Inner[Times, Clist /. {z[2] -> 1, z[3] -> 1, z[4] -> 1}, Clist, Plus];
Inlist = Inner1 /. Plus -> List /. {z[2] -> r[2]^2, z[3] -> r[3]^2, 
z[4] -> r[4]^2};
I1 = Integrate[ Inlist*r[2]*r[3]*Exp[-r[3]^2/(2 l[0]^2)], {r[3], 0, Infinity},
Assumptions -> (l[0] > 0)];
I2 = Integrate[ I1*Exp[-r[2]^2/(2 l[0]^2)], {r[2], 0, Infinity},
Assumptions -> (l[0] > 0)];
I3 = Integrate[ I2*r[4]*Exp[-r[4]^2/(2 l[0]^2)], {r[4], 0, Infinity},
Assumptions -> (l[0] > 0)] /. Plus -> List;
Ke = 0;KeN = 0;KeM = 0;CoefPoly1 = CoefPoly + 1;
Do[
I5 = Times[I3, i];
Ke = Ke + I5;
I6 = Integrate[ r[1]^(2 i - 1)*Exp[-r[1]^2/(2 l[0]^2)], {r[1], 0, Infinity},
Assumptions -> (l[0] > 0)];
KeN = KeN + I6;
KeM = KeM + FunctionExpand[(2 l[0]^2)^i*MeijerG[{{1}, {1}}, {{1/2, i}, {-1/2}}, 
Ne*Nu]/4],{i, 1, CoefPoly1}];
NomiIntegralr1 = Reverse[Plus @@ Ke /. Plus -> List];
KeN = Plus @@ Flatten[KeN] /. Plus -> List;
KeM = N[Plus @@ Flatten[KeM]] /. Plus -> List;
SumNomi = N[Plus @@ Flatten[ Inner[Times, NomiIntegralr1, KeM, Plus]]];
SumDenomi = N[Flatten[ Inner[Times, NomiIntegralr1, KeN, Plus]]];
SumGloVeb = Divide[SumNomi, SumDenomi];
EnergyVeb = N[Times[SumGloVeb, (-2*Rho*SN/R) e[0]^2], 8]; 
strm = OpenWrite["EnergyVeb"];
Write[strm, EnergyVeb];
Close[strm];
Print[EnergyVeb];] // Timing -2.25835527 e[0]^2/l[0].
\end{verbatim}

\ukrainianpart

\title{Аналітичні результати збуджених електронних станів при $\upsilon=1/3$ та методи   Лафліна-Джейна мікроскопічної хвильової функції}
\author{M.A. Aммар}
\address{Лабораторія фізики експериментальних методів і їх застосувань, Університет Медеї, Алжир}

\makeukrtitle

\begin{abstract}
	У даній роботі досліджено властивості двомірного електронного газу, який є під дією  сильного магнітного поля і охолоджений при низькій температурі. Подано точні аналітичні результати для енергій в основному стані. Ці
	результати стосуються систем аж до $ N_ {e} $ = 10 електронів, обчислені в режимі 
	цілочисельного квантового ефекта Гола (IQHE) при коефіцієнті заповнення 
	$\upsilon$ =1. Для здійснення обчислень  використано метод складних полярних
	координат. Слід зауважити, що система електронів у режимі квантового ефекту Гола, значною мірою спирається  на геометрію диска для скінчених систем електронів
	з довільними значеннями $ N_ {e} $ від 2 до 10 частинок. Результати, 
	отримані з допомогою аналітичних обчислень, добре узгоджуються з результатами Чіфт'я [Ciftja O., J. Math. Phys., 2011, \textbf{52}, 122105], де отримано представлення для  інтегралів добутків функцій Бесселя. І нарешті,  досліджено енергiї композитних ферміонів  збуджених станів для декількох систем при $\upsilon $ =1/3 та відповідність між дробовим квантовим ефектом Гола (FQHE) і цілочисельним квантовим ефектом Гола (IQHE).

	\keywords аналітичний метод, дробовий квантовий ефект Гола (FQHE), цілочисельний квантовий ефект Гола (IQHE), кулонівська взаємодія, квантовий ефект Гола, 2D електронний газ
\end{abstract}

\end{document}